\documentclass[useAMS,usenatbib, usegraphicx]{mn2e}
\usepackage{pslatex}

 \usepackage[dvips]{color}
\usepackage{color, colortbl}
\definecolor{LightCyan}{rgb}{0.5,1,1}

\newcommand{\kms}{\mbox{km}\,\mbox{s}^{-1}}
\newcommand{\target}{KOI-74}

\title[]{Mass ratio from Doppler beaming and R\o mer delay versus ellipsoidal modulation in the Kepler data of \target\thanks{Partly based on observations made with the Mercator Telescope, operated by the Flemish Community, and the William Herschel Telescope, both located at the Spanish Observatorio del Roque de los Muchachos of the Instituto de Astrof\'isica de Canarias.}}

\author[S. Bloemen et al.]{S. Bloemen$^{1,2}$\thanks{E-mail: steven.bloemen@ster.kuleuven.be}, T. R. Marsh$^{3}$, P. Degroote$^{1,2}$,  R. H. \O stensen$^{1}$,  P. I. P\'apics$^{1}$,  \newauthor C. Aerts$^{1,2,4}$, D. Koester$^{5}$, 
B. T. G\"ansicke$^{3}$, E. Breedt$^{3}$,  R. Lombaert$^{1}$, S. Pyrzas$^{3}$,   \newauthor  C. M. Copperwheat$^{3}$,  K. Exter$^{1}$, 
G. Raskin$^{1}$, H. Van Winckel$^{1}$, S. Prins$^{1}$,   W. Pessemier$^{1}$, \newauthor
Y. Fr\'emat$^{6}$,  H. Hensberge$^{6}$, 
A. Jorissen$^{7}$, S. Van Eck$^{7}$
\\
$^{1}$Instituut voor Sterrenkunde, Katholieke Universiteit Leuven, Celestijnenlaan 200 D, B-3001 Leuven, Belgium\\
$^{2}$Kavli Institute for Theoretical Physics, Kohn Hall, University of California, Santa Barbara, CA 93106, USA\\
$^{3}$Department of Physics, University of Warwick, Coventry CV4 7AL, UK\\
$^{4}$Department of Astrophysics, IMAPP, Radboud University Nijmegen, PO Box 9010, NL-6500 GL Nijmegen, the Netherlands\\
$^{5}$Institut f\"ur Theoretische Physik und Astrophysik, University of Kiel, D-24098 Kiel, Germany\\
$^{6}$Royal Observatory of Belgium, 3 Avenue circulaire, B-1180 Brussels, Belgium\\
$^{7}$Institut d'Astronomie et d'Astrophysique, Universit\'e Libre de Bruxelles, CP 226, Boulevard du Triomphe, B-1050 Brussels, Belgium
}
\begin{document}
\hyphenation{meas-ured HERMES at-mos-phere}
\date{Accepted 2012 February 24.  Received 2012 February 24; in original form 2012 January 22.}

\pagerange{\pageref{firstpage}--\pageref{lastpage}} \pubyear{2011}

\maketitle

\label{firstpage}

\begin{abstract}
We present a light curve analysis and radial velocity study of \target, an eclipsing A star + white dwarf binary with a 5.2 day orbit. Aside from new spectroscopy covering the orbit of the system, we used 212 days of publicly available {\em Kepler} observations and present the first complete light curve fitting to these data, modelling the eclipses and transits, ellipsoidal modulation, reflection, and Doppler beaming. Markov Chain Monte Carlo simulations are used to determine the system parameters and uncertainty estimates. Our results are in agreement with earlier studies, except that we find an inclination of {$87.0\pm0.4^\circ$}, which is significantly lower than the previously published value. The altered inclination leads to different values for the relative radii of the two stars and therefore also the mass ratio deduced from the ellipsoidal modulations seen in this system. We find that the mass ratio derived from the radial velocity amplitude ({$q=0.104\pm0.004$}) disagrees with that derived from the ellipsoidal modulation ({$q=0.052\pm0.004$} assuming corotation). This was found before, but with our smaller inclination, the discrepancy is even larger than previously reported. Accounting for the rapid rotation of the A-star, instead of assuming corotation with the binary orbit, is found to increase the discrepancy even further by lowering the mass ratio to {$q=0.047\pm0.004$}. These results indicate that one has to be extremely careful in using the amplitude of an ellipsoidal modulation signal in a close binary to determine the mass ratio, when a proof of corotation is not firmly established. The same problem could arise whenever an ellipsoidal modulation amplitude is used to derive the mass of a planet orbiting a host star that is not in corotation with the planet's orbit.

The radial velocities that can be inferred from the detected Doppler beaming in the light curve are found to be in agreement with our spectroscopic radial velocity determination. We also report the first measurement of R\o mer delay in a light curve of a compact binary. This delay amounts to {$-56\pm17$}\,s and is consistent with the mass ratio derived from the radial velocity amplitude. The firm establishment of this mass ratio at {$q=0.104\pm0.004$} leaves little doubt that the companion of \target\ is a low mass white dwarf.
\end{abstract}


%%%%%%%%%%%%%%%%%%%%%%%%%%%%%%%%%%%%%%%%%%%%
%%  INTRODUCTION
%%%%%%%%%%%%%%%%%%%%%%%%%%%%%%%%%%%%%%%%%%%%

\section[]{Introduction}\label{sec_intro}
The primary science goal of the {\em Kepler} Mission is the detection of Earth-like exoplanets, but its highly accurate photometric observations also reveal hundreds of eclipsing binary stars \citep{SlawsonPrsa2011} and are well suited for the study of stellar variability at unprecedentedly low amplitudes \citep{DebosscherBlomme2011}. In this paper, we present an analysis of 212\,d of {\em Kepler} data of the eclipsing binary \target\  (KIC 6889235), and spectra taken at different orbital phases. 

The system consists of a main-sequence A star primary and a less massive companion. Its light curve shows deeper eclipses than transits\footnote{We use the terms `transit' and `eclipse' to indicate, respectively, the occultation of the A-star by the compact object and the occultation of the compact object by the A-star half an orbit later.}, which implies that the companion is hotter than the primary. There is a clear asymmetric ellipsoidal modulation pattern in which the flux maximum after the transits is larger than the maximum after the eclipses.  \citet{RoweBorucki2010}  suggested that the asymmetry in the ellipsoidal modulation be due to a star spot. Van Kerkwijk et al.\ (2010) instead attributed it to Doppler beaming. This effect is caused by the stars' radial velocities which shift the spectrum, modulate the photon emission rate and beam the photons somewhat in the direction of motion. The effect was, as far as we are aware, first discussed in \cite{ShakuraPostnov1987} and first observed by \cite{MaxtedMarsh2000}. Its expected detection in Kepler light curves was suggested and discussed by \cite{LoebGaudi2003} and \cite{ZuckerMazeh2007}. The detection of Doppler beaming in Kepler light curves has led to the discovery of several non-eclipsing short-period binary systems \citep{FaiglerMazeh2011} and it has been shown that it can also be observed in planetary systems \citep[see e.g.][]{MazehFaigler2010,ShporerJenkins2011}.  \citet{BloemenMarsh2011} detected Doppler beaming in the {\em Kepler} light curve of KPD\,1946+4340 and presented the first comparison between a radial velocity amplitude derived from Doppler beaming with a spectroscopic value. In the case of KPD\,1946+4340, the results were found to be consistent. For KOI-74, spectroscopic radial velocity measurements were recently presented by \cite{EhrenreichLagrange2011}. In this paper, we present independent spectroscopic radial velocity measurements which we compare with the photometric radial velocity amplitude prediction. 

Earlier analyses of the Kepler light curve of \target\  are presented in \citet{RoweBorucki2010} and \citet{van-KerkwijkRappaport2010}. \citet{RoweBorucki2010} measured the mass ratio of the system from the amplitude of the ellipsoidal modulation and found a companion mass of 0.02-0.11\,M$_\odot$. Van Kerkwijk et al.\ (2010) built on these results (e.g.\ they used the inclination value report by \citealt{RoweBorucki2010}) but claimed that the mass ratio of the system could not be determined reliably from the ellipsoidal modulation amplitude. Instead, they used the radial velocity information from the Doppler beaming signal to derive a companion mass of $0.22\pm0.03$\,M$_\odot$. They concluded that the companion has to be a low mass white dwarf and showed that the system properties are in good agreement with a binary that has undergone a phase of stable 
Roche lobe overflow from the more massive star to the less massive star. {This puts \target\ in an evolutionary stage that follows on that of systems such as  WASP J0247-2515, which \citet{MaxtedAnderson2011} recently identified as a binary consisting of an A-star and a red giant core stripped from its envelope.} Up to now, 4 close binaries consisting of a white dwarf and a main sequence star of spectral type A or F have been found in \emph{Kepler} data \citep{RoweBorucki2010, van-KerkwijkRappaport2010,CarterRappaport2011,BretonRappaport2011}.

We remodel the Kepler light curve of \target, adding an additional 175\,d of data compared to the previous studies and perform Markov Chain Monte Carlo (MCMC) simulations to explore the uncertainty on the derived system parameters. We will revisit the issue of the true mass ratio of the system, using input from the light curve analysis (Doppler beaming, ellipsoidal modulation and R\o mer delay) and spectroscopy.

%%%%%%%%%%%%%%%%%%%%%%%%%%%%%%%%%%%%%%%%%%%%
%%  Spectroscopy
%%%%%%%%%%%%%%%%%%%%%%%%%%%%%%%%%%%%%%%%%%%%

\section[]{Spectroscopy} \label{sec_spec}

\begin{figure}
\includegraphics[width=84mm]{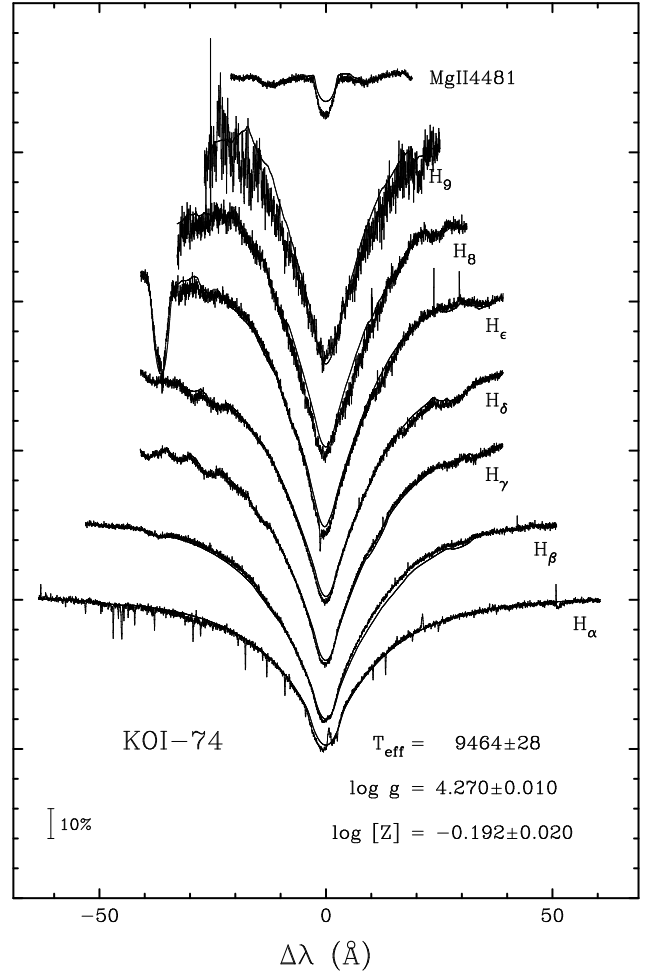}
 \caption{Fit to the spectral lines of \target\ using the average HERMES spectrum (after shifting the spectra to the rest frame of the primary) and the solar metallicity synthetic spectra of \citet{MunariSordo2005}. The uncertainties on the parameters indicated on the figure only reflect the formal errors on the fit.}
  \label{FIG_specfit}
\end{figure}

\begin{figure}
\includegraphics[width=84mm]{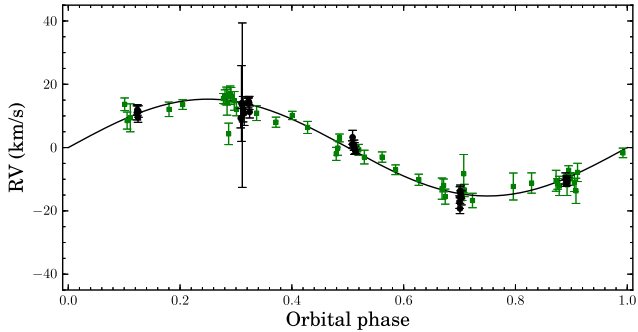}
 \caption{Radial velocity curve of the primary of \target, {measured by fitting a Gaussian profile to the mean profile of three Balmer lines (see text for details).} The radial velocity measurements from ISIS@WHT spectra are represented by black circles, the measurements from HERMES@Mercator spectra by green squares, folded on the orbital period. The systemic velocity derived from both datasets has been subtracted. {The error bars are scaled to deliver a unit $\chi^2$ per degree of freedom.}}
  \label{FIG_RV}
\end{figure} 

To determine the spectral type, the rotational velocity and the radial velocity amplitude of the primary, we obtained 46 high resolution ($R\sim 85000$) \'echelle spectra using the HERMES spectrograph \citep{RaskinVan-Winckel2011} at the 1.2-m Mercator telescope (La Palma, Canary Islands). These were reduced using the standard instrument specific data reduction pipeline. Additionally, we obtained 29 spectra with the ISIS spectrograph mounted on the 4.2-m William Herschel Telescope (La Palma, Canary Islands). 
We averaged the HERMES spectra, after shifting out the primary's radial velocity. From the average spectrum, we measured the rotational velocity of the primary $v \sin i = 164 \pm 9\,\kms$, using the Fourier method presented in \citet{Gray1992}, from the Mg II $\lambda 4481$ line. {Compared to other techniques such as fitting broadened synthetic spectra, the Fourier method has the advantage of being rather unsensitive to other line broadening mechanisms \citep[see e.g.][]{Simon-DiazHerrero2007}. }The Mg II $\lambda 4481$ line is a doublet with components at 4481.126\,\AA\, and 4481.325\,\AA, which leads to an overestimation of $v \sin i$ \citep[see e.g.][]{RoyerGerbaldi2002}. By applying the same method to a synthetic spectrum with a comparable rotational broadening, we find that we can expect our $v \sin i$ measurement to be overestimated by about $10\,\kms$ due to the double nature of the line. Given this overestimation, our result is in line with the values mentioned in the {\em Note in proof} of \citet{van-KerkwijkRappaport2010}, $150\,\kms$, and in \cite{EhrenreichLagrange2011}, $145 \pm 5\,\kms$. If the primary were in corotation with the binary orbit, we would expect $v \sin i \sim 25\,\kms$. The rotational velocity measured from spectroscopy thus implies that the primary is not in corotation but is instead a fast rotator. We have adopted $v \sin i = 150 \pm 10\,\kms$ for the analysis presented in this paper.
We also performed a spectral fit to the Balmer lines and Mg II $\lambda 4481$ using the solar metallicity synthetic spectra of \citet{MunariSordo2005} and assuming a rotational velocity of $150\,\kms$. The fits to the spectral lines are shown in Fig.~\ref{FIG_specfit}. {The small emission feature in the core of the H$\alpha$ and H$\beta$ lines does not originate from any of the two binary components but is caused by sky emission \cite[see also][]{EhrenreichLagrange2011}. It is absent in the ISIS spectra, for which we could perform a background subtraction during the data reduction.} We find that the primary has a gravity of $\log g \sim 4.27$ and an effective temperature of $T_{\rm{eff}} \sim 9\,500$K. These results are in  agreement with the spectral type A1V and $T_{\rm{eff}}=9\,400\pm150\,$K derived by \citet{RoweBorucki2010}{, from which they inferred a primary mass of $M_1=2.2 \pm 0.2\,{\rm M}_\odot$. We adopted $T_{\rm{eff}}=9\,500\pm250\,$K for our analysis.}

{We have used two techniques to derive the radial velocity amplitude of the A-star. The only metal line that is clearly detected in both the HERMES and ISIS spectra, is the Mg II line at $4481\,$\AA. We measured the radial velocities from that line by fitting a Gaussian profile to it. We also measured the radial velocities from the Balmer lines at $4102$\AA, $4340$\AA\ and $4861$\AA\ simultaneously, by first normalising the spectra using low order splines \citep[as described in][]{PapicsBriquet2012}, and then fitting a Gaussian to the core of the mean profile obtained by Least Squares Deconvolution \citep{DonatiSemel1997}. The measured radial velocity amplitudes and systemic velocities, assuming a circular orbit, are given in Table \ref{tab_RVs}. The uncertainties on individual data points have been scaled to deliver a unit $\chi^2$ per degree of freedom. The initial $\chi^2$, before rescaling the uncertainties, is given in the table. 
The radial velocities measured from the Balmer lines are shown in Fig.~\ref{FIG_RV}, folded on the orbital period using the {\em Kepler} ephemeris as given in \citet{RoweBorucki2010}. Some spectra were taken in bad seeing conditions, which is reflected by the large error bars on a few of the data points. We have adopted the weighted mean velocity amplitude of our measurements, $K_1=15.4 \pm 0.3\,\kms$, for the analysis presented in this paper. The weighted mean systemic velocity, is $\gamma=49.1\pm0.2\,\kms$. Our radial velocity amplitudes are in agreement with (but slightly lower than) \cite{EhrenreichLagrange2011}'s result of $18.2 \pm 1.7\,\kms$. We provide the radial velocities we have measured from our spectra in electronic form with this paper.}

\begin{table}
 \caption{{Radial velocity amplitudes ($K_1$) and systemic velocities ($\gamma$) of \target, measured from ISIS@WHT and HERMES@Mercator spectra. The uncertainties that are given are scaled to get a unit $\chi^2$ per degree of freedom. Initial reduced $\chi^2$ is given in the last column. Two techniques have been used: a Gaussian fit to the Mg II line at $4481\,$\AA and a Gaussian fit to the mean profile of the Balmer lines at $4102$\AA, $4340$\AA\ and $4861$\AA (see text for details).}}
 \label{tab_RVs}
 \begin{center}
 \begin{tabular}{llccc}
  \hline
Instrument  & Line(s) & $K_1\ (\kms)$ & $\gamma\ (\kms)$ &$\chi^2_{\rm init, reduced}$\\
  \hline
ISIS & Balmer & $15.8\pm0.4$ & $-50.9\pm0.3$&1.2\\
 & Mg II & $15.4\pm0.7$ & $-49.3\pm0.5$&1.6\\
HERMES & Balmer & $14.9\pm0.4$ & $-47.1\pm0.3$&0.5\\
 & Mg II & $16.5\pm1.3$ & $-51.0\pm1.0$&2.3\\
 \\
 \multicolumn{2}{l}{Weighted mean (adopted)} & $15.4 \pm 0.3 $ & $-49.1\pm0.2$ &\\
 
 \hline

  \end{tabular} \end{center}
\end{table}

%%%%%%%%%%%%%%%%%%%%%%%%%%%%%%%%%%%%%%%%%%%%
%%  LC MODEL
%%%%%%%%%%%%%%%%%%%%%%%%%%%%%%%%%%%%%%%%%%%%

\section[]{{\em Kepler} photometry} \label{sec_phot}
\begin{figure*}
\includegraphics[width=170mm]{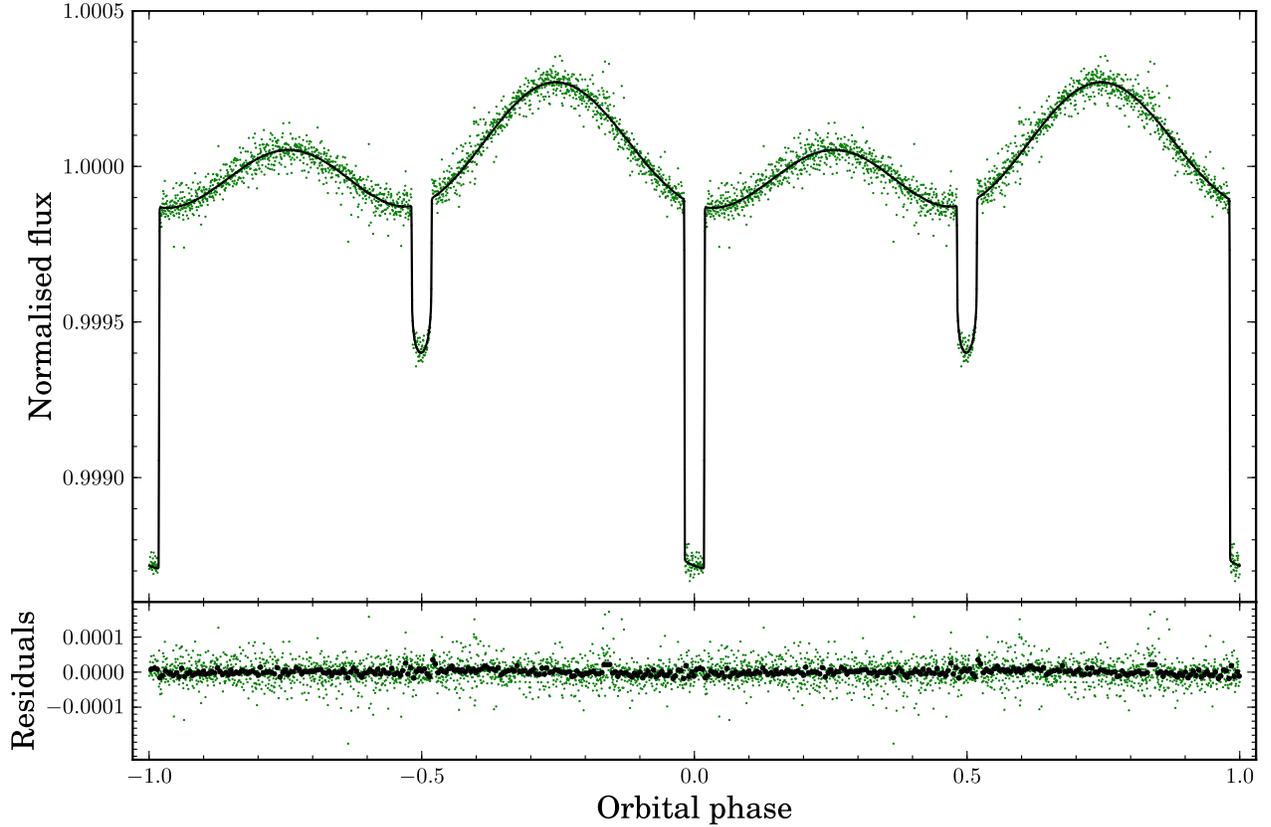}
 \caption{The {\em Kepler} Q0, Q1, Q2 and Q3 data of \target, folded on the orbital period and binned into 2000 phase bins. The black curve shows a typical model fit to the data. The residuals of the data points after subtracting the model are shown in the bottom panel, binned into 2000 phase bins (green) and 200 phase bins (black).}
  \label{FIG_LC}
\end{figure*}

The {\em Kepler} data from Q0 (quarter 0), Q1, Q2 and Q3 were retrieved from the public archive\footnote{Publicly released {\em Kepler} data can be downloaded from http://archive.stsci.edu/kepler/}. The data span 229\,d, resulting in a dataset of 212\,d of observations excluding the gaps. The data of Q0, Q1 and the first two months of Q2 are long cadence data (30\,m integrations), the last 27\,d of Q2 and the entire Q3 dataset (86\,d) are taken in short cadence mode (1\,m integrations).  The light curve (see Fig. \ref{FIG_LC} in this paper for a phase folded version) shows clear eclipses, transits, ellipsoidal modulation and Doppler beaming. \cite{RoweBorucki2010} and \cite{van-KerkwijkRappaport2010} already presented models for the light curve. We remodelled the light curve with the \texttt{LCURVE} code written by TRM, using more data, and find a lower orbital inclination than \cite{RoweBorucki2010} ($i=88.8 \pm 0.5^\circ$). This finding also has consequences for some of the system parameters derived by \cite{van-KerkwijkRappaport2010}, as they adopted the inclination of \cite{RoweBorucki2010} for their analysis. 

Below we discuss the various steps of our binary light curve modelling.

% -------------------------------------------------------------------
%  LC detrend
% -------------------------------------------------------------------

\subsection{Detrending of the Kepler data}\label{sec_LCM_detrend}
We extracted the Kepler data from the pixel data, rather than using the pipeline output. We first normalised the data by dividing out an initial model fit, to be able to determine the trends in the data. Next, we detrended the data by dividing out exponentials fitted to the two obvious instrumental decays during Q2 and Q3. We removed 101 long cadence points from the Q2 data, as well as 1299 short cadence datapoints from the Q3 data, which were too severely affected by instrumental effects to allow for accurate detrending.  We then divided out second order polynomials fitted to each segment of the light curve. While fitting the polynomials, we flagged 184 points as $>4$-$\sigma$ outliers, in an iterative fashion, leaving 166\,836 datapoints. We then folded the initial orbital fit back in. Finally, we binned the short cadence data out of eclipse to long cadence data since there are no binary signatures to be expected outside eclipse that require the time precision offered by short cadence. This way, the number of datapoints is reduced to 28\,086. 

Initial MCMC runs using this dataset (see Section \ref{sec_LCM_model}) showed, as expected, that there is no correlation between the adopted ephemerides ($T_0$ and $P$) and the other free parameters. We therefore decided to create a phasefolded version of the four months of short cadance data to speed up the simulations. We rebinned the light curve into 1800\,s bins out of eclipse and 30\,s bins in eclipse, to be left with 1717 data points.

We assume that there is no background contamination in the \emph{Kepler} data. We provide our final datasets in electronic form with this paper.

% -------------------------------------------------------------------
%  LC info
% -------------------------------------------------------------------

\subsection{What can we learn from the {\em Kepler} light curve?}\label{sec_analytic}

Fitting a light curve of a binary that shows a variety of effects (such as eclipses, ellipsoidal modulation and Doppler beaming) allows one to gain a lot of information on the binary's parameters. We first explore the potential before we discuss the actual light curve fits in Section~\ref{sec_LCM_model}.

A first useful constraint comes from the measured fractional transit depth, which reveals the ratio of the radii of the two stars: \begin{equation}\Delta F_{transit}/F = \left(R_2/R_1\right)^2.\end{equation} 

Given $R_2/R_1$ from the transit depth and $T_1$ from the spectral fit (Section~\ref{sec_spec}), the eclipse depth provides information about $T_2$. Using blackbody approximations and neglecting the contribution of the companion out of eclipse, one finds: \begin{equation}\frac{\Delta F_{eclipse}}{F} \approx \frac{F_2}{F_1} = \frac{\exp\left(h\nu/kT_1\right)-1}{\exp\left(h\nu/kT_2\right)-1}\left(\frac{R_2}{R_1}\right)^2. \end{equation}

The eclipse duration $t_e$ gives $R_1/a$ as a function of the inclination $i$ (assuming spherical stars): \begin{equation}(R_1/a)^2= \sin^2(\pi t_e/P)\, \sin^2 i  + \cos^2 i \label{eq_te} \end{equation} \citep{Russell1912} with P the orbital period and $a$ the separation between the two binary components.

In case of \emph{corotation} of the star with the binary orbit, the semi-amplitude of the ellipsoidal modulation can be approximated as\footnote{Van Kerkwijk et al.\ (2010) use a similar equation but with a $\sin^3 i$ term, which should be $\sin^2 i$.} \begin{equation} \frac{\Delta F_{ellipsoidal}}{F}= 0.15 \frac{\left(15+u_1\right)\left(1+\tau_1\right)}{3-u_1} \left(\frac{R_1}{a}\right)^3 q\,\sin^2 i \label{eq_qiEM}\end{equation} \citep{MorrisNaftilan1993, ZuckerMazeh2007} in which $q=M_2/M_1$ is the mass ratio of the two binary components, $u_1$ the primary's linear limb darkening coefficient and $\tau_1$ its gravity darkening coefficient.

With $R_1/a$ determined from the eclipse duration (Eq.~\ref{eq_te}), $q$ is then known as a function of $i$. In the case of \target, the primary is not in corotation but is a rapid rotator, as already suggested by \citet{van-KerkwijkRappaport2010} and confirmed by all measurements of the rotational velocity (see Section~\ref{sec_spec}). Van Kerkwijk et al.\ (2010) argue that rapid rotation has a very significant impact on the ellipsoidal modulation amplitude. \cite{Kruszewski1963} presented an expression for the Roche lobe potentials that accounts for the effects of asynchronous rotation. Although we will use this treatment in our light curve models, we will be careful when interpreting the mass ratio derived from our light curve modelling effort presented in Section~\ref{sec_LCM_model}. Our modelling setup is such that the mass ratio is the only parameter that is only constrained by the ellipsoidal modulation amplitude. Therefore, if the expression for the ellipsoidal modulation is invalid (e.g. because it assumes instantaneous adjustment of the star's surface to the ever-changing potential in the asynchronous case), this will manifest itself in the mass ratio being off.

The detection of Doppler beaming in the extremely accurate {\em Kepler} light curve \citep[see][]{van-KerkwijkRappaport2010} gives access to a second, independent, way to determine $q$ as a function of $i$. For velocities much lower than the speed of light, the Doppler beaming amplitude is proportional to the radial velocity amplitude of the A-star: 
\begin{equation}F_\lambda = F_{0,\lambda} \left( 1 - B \frac{v_r}{c}\right) \label{eq_DB} \end{equation}
(Eq.~2 in \citealt{BloemenMarsh2011}). The beaming factor \begin{equation}B = 5+ \rm{d}\ln F_\lambda / \rm{d} \ln \lambda \label{eq_BF}\end{equation} \citep{LoebGaudi2003} depends on the wavelength $\lambda$ of the observation and on the spectrum of the star. We determine this factor for the case of \target\ in Section~\ref{sec_coeffs}.

With $M_1$ given by the spectral type derived from spectroscopy (Section~\ref{sec_spec}), and $K_1$ derived from the Doppler beaming amplitude (Eq.~\ref{eq_DB}) or spectroscopy (Section~\ref{sec_spec}), we get the following relation where $q$ and $i$ are the only unknowns: \begin{equation} K_1^3 = \frac{q^3}{\left( 1+q \right)^2} \frac{2\pi G M_1}{P} \sin^3 i . \label{eq_qiRV}\end{equation}
This equation allows one to derive the mass ratio without relying on the amplitude of the ellipsoidal modulation (as long as the deformation of the primary is small enough not to affect the spectroscopic parameter determinations). The mass ratio $q$ derived from the radial velocity amplitude using Eq.~\ref{eq_qiRV} is not influenced by the rapid rotation, contrary to the mass ratio that can be derived from the ellipsoidal modulation amplitude via Eq.~\ref{eq_qiEM}. 

Recently, \citet{Kaplan2010} showed that one can use an effect similar to the R\o mer delay to derive the radial velocity amplitude of the secondary, in double white dwarf binaries with circular orbits and mass ratios significantly different from 1. In such systems, a light travel time difference causes the time between the primary and secondary eclipses to be different from $P/2$ by
\begin{equation}\delta t = \left( K_2-K_1 \right)\frac{P}{\pi c}  \label{eq_Roemer} \end{equation}
in which $c$ denotes the speed of light \cite[see][]{Kaplan2010}.  After substitution of $q=\frac{K_1}{K_2}$, we find the following expression: \begin{equation} q=\left( 1- \frac{\pi c\, \Delta t }{P K_1}\right)^{-1}.\end{equation} If we can measure this R\o mer delay, it would allow us to derive the mass ratio in yet another way, this time independently from the mass of the primary which was estimated from spectral analysis. As far as we are aware, this technique has not yet been applied to any system in practice. \target\ is a good candidate to put the theory to the test, since the expected time difference is of the order of a minute. 

% -------------------------------------------------------------------
%  GDC, LDC, BF
% -------------------------------------------------------------------

\subsection{Gravity darkening, limb darkening and Doppler beaming coefficients}\label{sec_coeffs}
The gravity darkening coefficient (see e.g.\ \citealt{Claret2003}) of the primary was calculated by integrating ATLAS model spectra (\citealt{CastelliKurucz2004}) over the {\em Kepler} bandpass. We took into account the estimated reddening of $E(B-V)=0.15$ (\emph{Kepler} Input Catalog) by reddening the model spectra following \cite{CardelliClayton1989}. {Reddening marginally influences the gravity darkening coefficient because it is bandpass dependent.} Assuming solar metallicity, $v_{{\rm turb}}=2$ km/s, $T_{{\rm eff}}=9500 \pm 250\,$K and $\log(g)=4.3\pm0.1$, and using Eq.~1 from \cite{BloemenMarsh2011}, we found the gravity darkening coefficient to be $\beta_K = 0.55 \pm 0.05$. {We have used a physical gravity darkening coefficient of $\mathrm{d} \log T/\mathrm{d}\log g=0.25$.}

Using the same assumptions, we computed limb darkening coefficients for the A-star. We adopted the 4-parameter limb darkening relation of \citet[equation 5]{Claret2004LDC} with $a_1=0.576$, $a_2=0.118$, $a_3=-0.039$ and $a_4=-0.016$. For the white dwarf companion, we used a model atmosphere for a DA white dwarf with $T_{\rm{eff}}=13\,000$K and $\log g=6.5$ \citep{Koester2010} and found  $a_1=0.372$, $a_2=0.518$, $a_3=-0.540$ and $a_4=0.178$.

For the broadband {\em Kepler} photometry, the wavelength-specific beaming factor ($B$ in Eq.\,\ref{eq_DB}) has to be replaced by a bandpass-integrated photon weighted beaming factor 
\begin{equation}
\left<B\right> = \frac{\int
  \epsilon_\lambda \lambda F_\lambda B \, d\lambda}{\int \epsilon_\lambda
  \lambda F_\lambda \, d\lambda}
  \end{equation}
 (Eq.~3 in \citealt{BloemenMarsh2011}) in which $\epsilon_\lambda$ is the response function of the {\em Kepler} bandpass and $B$ the monochromatic beaming factor (Eq.~\ref{eq_BF}). Taking reddening into account, the beaming factor is found to be $\left<B\right>= 2.19 \pm 0.04$, which compares well with \cite{van-KerkwijkRappaport2010}'s value of 2.21. For an unreddened spectrum, the beaming factor would be $\left<B\right>= 2.22 \pm 0.04$.

% -------------------------------------------------------------------
%  Model and parameter uncertainties
% -------------------------------------------------------------------

\subsection{Modelling code and MCMC setup} \label{sec_LCM_model}
Our light curve modelling code, \texttt{LCURVE} \citep[for a description of the code, see][Appendix A]{CopperwheatMarsh2010}, uses grids of points on the two binary components to calculate the total flux that is visible from the system at different orbital phases. It accounts for Doppler beaming, eclipses and transits, ellipsoidal modulation and reflection effects. The code can also account for lensing effects \cite[implemented following][]{Marsh2001}, which occur when the white dwarf transits the primary. In the models of KOI-74 presented here, however, we did not include these lensing effects. The white dwarf radius we would find by including lensing would be slightly larger but we estimate the difference at about one percent only, which is far lower than the uncertainty on the parameter. In addition, estimation of the lensing requires knowledge of the mass of the white dwarf, which owing to the difficulty in deducing the mass ratio from the ellipsoidal variations, was not easily calculated during the MCMC runs. 

\begin{figure}
\includegraphics[angle=-90, viewport=0 0 420 430, width=84mm]{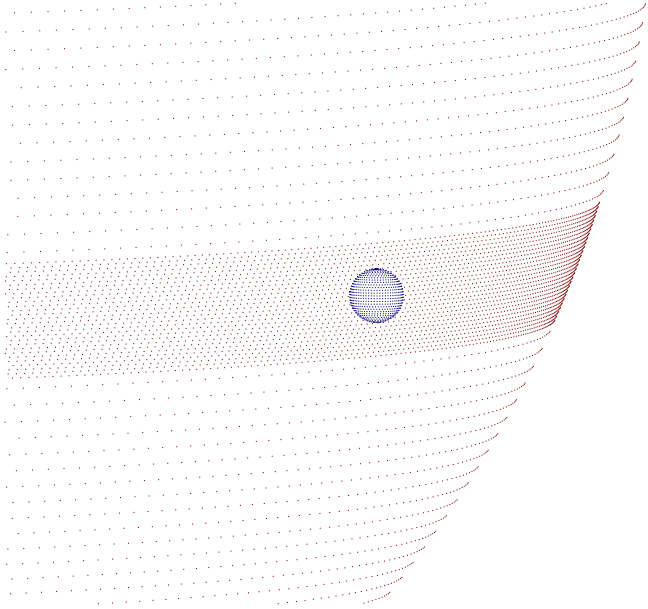}
 \caption{Illustration of the grids of points we used on the two binary components to model the light curve of KOI-74 (only about one point in 4   is shown on the A-star, and about one in 10 on the white dwarf). To achieve a high enough numerical precision, at the transit phases, a denser strip is used on the primary at the location where it gets occulted by the white dwarf. The system is shown with an inclination of 86.4$^\circ$, at orbital phase 0.02. }
  \label{FIG_grids}
\end{figure}

A typical light curve fit is shown in Fig.\ \ref{FIG_LC}, together with the phase folded light curve and the residuals. Because of the large difference in the radii of the stars, it turned out to be difficult to get the numerical noise at a lower level than the scatter on the observational datapoints of the superb {\em Kepler} light curve. The grid on the A-star would have to consist of millions of points to get to the required accuracy level, which is particularly difficult during the transits. We therefore replaced the grid of uniformly distributed points, at the transit phases, by a grid with a denser strip on the A-star at the region of the star that gets occulted during the white dwarf transits. A graphical representation of the grid is shown in Fig.\ \ref{FIG_grids}. We used 92\,716 points (on 270 latitude strips) on the A-star outside the transits and 178\,468 points during the transits (adding 7 latitude and 4 longitude points per coarse grid point in the strip). The flux from the white dwarf was modelled with a 12\,724-points grid (100 latitude strips) at all orbital phases.

To account for a possible detection of the R\o mer delay, we introduced a $\delta t$ parameter and computed the light curve at phase $\phi'$ instead of the requested phase $\phi$, with
\begin{equation}\phi' = \phi + \frac{\delta t}{P}\left(\frac{\cos\left(2 \pi \phi\right)}{2}-0.5\right).\end{equation}
This essentially results in a maximum R\o mer delay of $-\delta t$ at the phase of the eclipse, while the phase of the transit is untouched. 

In an attempt to correctly model the ellipsoidal modulation taking into account the effects of rapid rotation, we implemented the adapted Roche lobe potentials as given by \citet{Kruszewski1963}. The modelling procedure, using Markov Chain Monte Carlo simulations, is identical to the one used in \citet[][Sections 3.3 and 3.4]{BloemenMarsh2011} for the modelling of the light curve of KPD\,1946+4340. In the initial runs using the full dataset, the orbital period ($P_{\rm{orb}}$), the time zero point ($T_0$), the inclination ($i$), the temperature of the white dwarf ($T_2$), the mass ratio ($q$), the beaming factor ($\left<B\right>$) and the R\o mer delay ($\delta t$) were kept as free parameters. The computation of one synthetic light curve at the 28\,086 time points of the dataset, for which we computed the light curve at about 127\,000 phases to be able to account for the finite integration time of the observations, took about 5 minutes of CPU time. Due to the strong degeneracy between the inclination and the mass ratio, the MCMC chains took too much time to sample the whole parameter space. We therefore fixed the orbital period ($P_{\rm{orb}}$) and the time zero point ($T_0$) at the optimal values of the initial runs, and used the phasefolded light curve (see Section \ref{sec_LCM_detrend} for details) for the final runs.

To account for the finite integration times of the observations, we oversampled our light curves in time space. The 30\,s phasebins during eclipses and transits were oversampled by a factor 5, and the 1800\,s phasebins out of eclipse by a factor 3. For the oversampling, we approximated the effective integration time for each bin by $\sqrt{\left(I\right)^2+\left(x\right)^2}$, in which $I$ is the integration time of the binned short cadence data points (about 58\,s) and $x$ is the width of the bin (30\,s or 1800\,s).
To explore the effects of rapid rotation, we ran MCMC chains assuming corotation (as in the case of KPD\,1946+4340), as well as chains treating the spin rate as a free parameter with $v \sin i=150\pm10\,\kms$ as a prior. {This prior was implemented as a constraint on $(2\pi R/P)\sin i$.} In total, we computed over {450\,000} light curves in the corotation chain, plus about {400\,000} light curves in the fast rotation chain with spectroscopic prior on the spin rate. {Every tenth model that was computed was stored and used to determine the system parameters.}

We set the limb darkening coefficients to the values found in Section~\ref{sec_coeffs} and used priors on the {flux weighted} temperature of the primary ($T_1= 9500\pm250\,$K, taken from spectroscopy, see Section~\ref{sec_spec}), the gravity darkening coefficient of the primary ($\beta_K = 0.55 \pm 0.05$, see Section~\ref{sec_coeffs}) and the radial velocity amplitude of the primary ({$K_1= 15.4 \pm 0.3\,\kms$}, taken from spectroscopy, see Section~\ref{sec_spec}). If the amplitude of the Doppler beaming effect is consistent with the radial velocity amplitude, as was the case for KPD\,1946+4340 \citep{BloemenMarsh2011}, the results of the analysis will be identical when the prior on $K_1$ would be replaced by a prior on $\left<B\right>$. In our approach, putting a prior on $K_1$ and treating $\left<B\right>$ as a free parameter, we can judge whether the Doppler beaming effect has the expected amplitude by comparing the $\left<B\right>$ that results from our MCMC analysis with the one found from atmosphere models in Section~\ref{sec_coeffs}.

% -------------------------------------------------------------------
%  Discussion
% -------------------------------------------------------------------

\subsection{MCMC results and discussion} \label{sec_LCM_disc}
{As can be seen in Fig.\ \ref{FIG_LC}, the light curve model fits the observed data very well. There is some structure in the residuals around the transit ingress and egress (around orbital phase 0.5), but it is barely significant. This structure can for example result from a small difference between the assumed limb darkening coefficients (which were derived for a spherical star) and the true limb darkening of the A-star.}
The parameters derived from our MCMC runs that account for the rapid rotation of the primary, using the prior on $v \sin i$, are summarized in Table \ref{tab_pars}. \cite{RoweBorucki2010} determined $M_1=2.2 \pm 0.2\,{\rm M}_\odot$ based on the spectral type. Using the evolutionary model grids of \cite{BriquetAerts2011} and our spectroscopic determinations of $T_{\rm eff}$ and $\log g$, we find the same result, which we adopted for our analysis. The results of the MCMC chains for the corotating case are nearly identical except for the derived mass ratios. The mass ratio from the models assuming corotation is also listed in the table, to allow for comparison with previously published values. The most important difference between our results and those previously published, is that our inclination {$i=87.0 \pm 0.4^\circ$} differs by $\sim 3\sigma$ from the value derived by \cite{RoweBorucki2010}, $i=88.8 \pm 0.5^\circ$. In that paper, the inclination was determined by fitting the eclipses and transits using the analytical formulae of \cite{MandelAgol2002}, which include limb darkening but not gravity darkening. Furthermore, it is possible that the authors did not account for the finite exposure times, which significantly smear out the eclipse ingresses and egresses in the long cadence data they had available. Fitting only the 43\,d of long cadence data that \cite{RoweBorucki2010} used, we find an uncertainty on the inclination of $0.9^\circ$, compared to their more optimistic value of $0.5^\circ$. The 4 months of short cadence observations allowed us to determine the inclination more reliably.

\begin{table}
 \caption{Properties of \target.\ The mass of the primary, the effective temperature and the rotational velocity are derived from spectroscopy. The other parameters are the result of our MCMC analysis of the {\em Kepler} data. The two different mass ratios are derived from the information contained in the ellipsoidal modulation amplitude ($q_{\rm{ell}}$) and the radial velocity ($q_{\rm{RV}}$). The mass of the secondary is derived using $q_{\rm{RV}}$.}
 \label{tab_pars}
 \begin{center}
 \begin{tabular}{lcc}
  \hline
  & Primary (A-star) & Secondary (WD) \\
  \hline
$P_{\rm{orb}}$ (d) &  \multicolumn{2}{c}{$5.188675(4)$} \\
%$T_0$ (BJD) &  \multicolumn{2}{c}{$2\,455\,023.XXXX(8)$} \\
$i$  (deg) & \multicolumn{2}{c}{$87.0 \pm 0.4$} \\
%$R/a$&  $0.XXX \pm 0.XXX$ & $0.XXX \pm 0.XXX$ \\
$T_{\rm{eff}}$ (K) &  $9500 \pm250$ & {$14$\,$500 \pm 500$}\\
$q_{\rm{ell,corot}}$ & \multicolumn{2}{c}{$ 0.052\pm 0.004$} \\
$q_{\rm{ell,vsini}}$ & \multicolumn{2}{c}{$0.047\pm 0.004$} \\
$q_{\rm{RV}}$ & \multicolumn{2}{c}{$0.104  \pm 0.004$} \\
$R$ $\left(\rm{R}_\odot\right)$&  {$2.14 \pm 0.08$} & {$0.044 \pm 0.002$} \\
$M$ $\left(\rm{M}_\odot\right)$&  $2.2 \pm 0.2$ & {$0.228 \pm 0.014$} \\
$v \sin i$ $\left( \kms \right)$&  $150 \pm 10$ & $--$ \\
$\delta t_{\rm{R\o mer}} (s)$ & \multicolumn{2}{c}{$-56 \pm 17$}\\
$\left< B \right> $ & {$2.24 \pm 0.05$} & $--$ \\
\hline

  \end{tabular} \end{center}
\end{table}

\begin{figure}
\includegraphics[width=84mm]{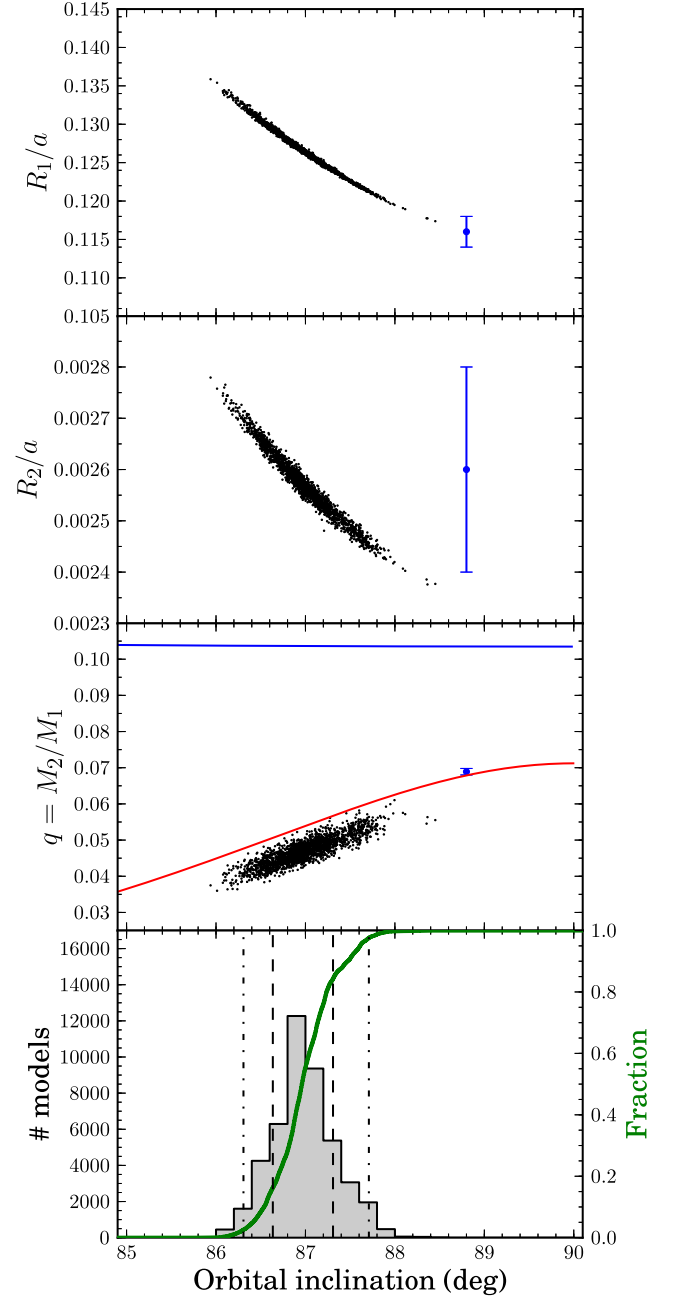}
 \caption{Illustration of the correlation between the inclination of the system and the mass ratio and radii of the stars. The bottom panel shows the distribution of the models that were accepted in our MCMC runs that account for the rapid rotation of the primary. The dashed (dot-dashed) line indicates the 68 (95) per cent confidence interval. The black dots represent a random selection of the models. The blue dots with error bars show the results from \citet{van-KerkwijkRappaport2010}. The red line on the mass ratio plot is the theoretical relation based on Eq.~\ref{eq_te} and Eq.~\ref{eq_qiEM}, assuming corotation. The blue line shows the mass ratio derived from $K_1$ (Eq.~\ref{eq_qiRV}).}
  \label{FIG_MCMC}
\end{figure}

The distribution of the inclination values in our MCMC runs using the prior on $v\sin i$ is shown in the bottom panel of Fig.\ \ref{FIG_MCMC}. The two top panels of the figure show our radius estimates (relative to the separation of the two stars, $a$) as a function of the inclination. The blue points with error bars indicate the values found by \cite{van-KerkwijkRappaport2010}, who did not derive the inclination independently but instead adopted the value of \cite{RoweBorucki2010}. Due to our lower preferred inclination, the radii inferred in our analysis are slightly larger than the ones obtained by \cite{van-KerkwijkRappaport2010}.

The third panel of Fig.\ \ref{FIG_MCMC} shows the mass ratio as a function of the inclination. The red line is the theoretical relation, assuming corotation, based on the constraints offered by the transit duration (Eq.~\ref{eq_te}) and the ellipsoidal modulation amplitude (Eq.~\ref{eq_qiEM}). The distribution of points from our corotation runs, falls nicely around this line (not shown). The blue line shows the mass ratio derived from the spectroscopic $K_1$, or equivalently, as we will see further in this Section, the Doppler beaming amplitude. The discrepancy between the mass ratios derived from the ellipsoidal modulation amplitude on one hand, and from the radial velocity information on the other hand, was already noted by \cite{van-KerkwijkRappaport2010}. The results of the models that account for rapid rotation of the primary are plotted with black dots. It is striking that these models result in lower rather than higher mass ratios, thus only increasing the discrepancy compared to the models assuming corotation. The lower preferred inclination value also contributes to the increase in the discrepancy between the two mass ratios, compared to the discrepancy shown by \cite{van-KerkwijkRappaport2010}. This can easily be understood: as $i$ gets lower, $R_1/a$ has to increase to make the model fit the observed transit duration (Eq.~\ref{eq_te}); but a higher $R_1/a$ would lead to a higher ellipsoidal modulation amplitude, which is then compensated in the simulations by lowering $q_{\rm ell}$ (Eq.~\ref{eq_qiEM}). 

The beaming factor derived from the MCMC chains, {$\left<B\right> = 2.24 \pm 0.05$}, agrees with the beaming factor derived from spectroscopy in Section~\ref{sec_coeffs} ($\left<B\right>= 2.19 \pm 0.04$, taking reddening into account). We could thus have modelled the binary equally well without spectroscopic radial velocity information, just relying on the Doppler beaming amplitude in the {\em Kepler} light curve, verifying \cite{van-KerkwijkRappaport2010}'s approach. 

One can argue that the mass ratios determined from the ellipsoidal modulation amplitude and radial velocity information can be brought into agreement by assuming a high contamination of the {\em Kepler} light curve by background stars, since this would increase the observed ellipsoidal modulation amplitude. The consistency between the observed Doppler beaming amplitude (which would also increase if one assumes a higher contamination) and the spectroscopic radial velocity amplitude, however, implies that our assumption of no background contamination in the Kepler light curve has to be correct up to a few per cent. 

\begin{figure}
\includegraphics[width=84mm]{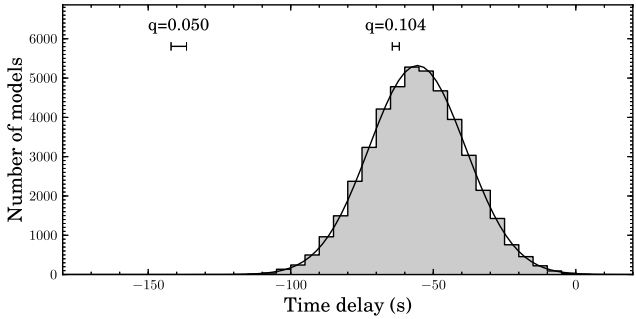}
 \caption{Distribution of the R\o mer delay as fitted by our MCMC runs. The delays expected for mass ratios of $q=0.050$ and $q=0.104$ are indicated for comparison.}
  \label{FIG_Roemer}
\end{figure}

We were also able to measure the R\o mer delay at {$\delta t=-56\pm17\,$s}, as can be seen on the distribution plot from our MCMC chains on Fig.~\ref{FIG_Roemer}. The expected R\o mer delay for a mass ratio of {$q=0.104$ is  $\delta t = -63.2 \pm 1.2$\,s}, while for $q=0.050$ we would expect {$\delta t = -139 \pm 3$\,s} (the error bars account for the uncertainty on $K_1$ measured from spectroscopy). The R\o mer delay estimates only depend on $P_{\rm{orb}}$, $K_1$ and the mass ratio. With $P_{\rm{orb}}$ and $K_1$ firmly established from both spectroscopy and photometry, the measurement of the delay proves that the true mass ratio of the system is the one derived from the radial velocity amplitude, under the condition that the orbit is circular. Note that, while $K_1$ can be derived from the Doppler beaming amplitude, a measurement of the R\o mer delay  allows one to also derive $K_2$ directly from the light curve if $M_1$ is known.

% -------------------------------------------------------------------
%  Discussion
% -------------------------------------------------------------------

\subsection{Variability in residuals} \label{sec_resid}

\begin{figure}
\includegraphics[width=84mm]{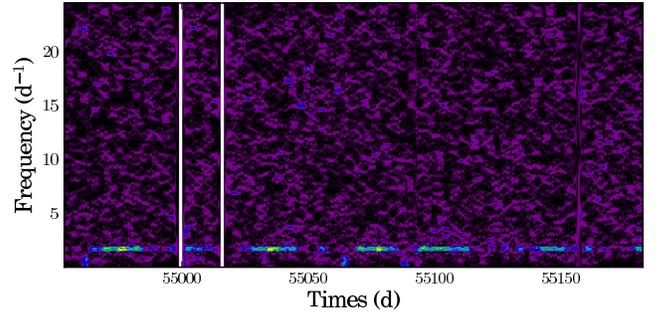}
 \caption{Time-frequency analysis of the residuals of the short cadence data after subtracting our best fitting binary light curve model. The observed variability at $\sim1.3$\,d$^{-1}$ might be related to the spin period of the primary star.}
  \label{FIG_timefreq}
\end{figure}

We performed a time-frequency analysis on the residuals of the short cadence data after subtracting our best model. The time-frequency diagram is shown in Fig.\ \ref{FIG_timefreq}. We detected significant variability  \citep[following the criteria given in][]{DegrooteAerts2009}, but all at low amplitudes ($\sim10$ parts per million). We found variability with periods of $\sim3\,$d, which is a known instrumental artefact \citep[see][]{ChristiansenVan-Cleve2011}. We also found variability with a period of $0.5918\pm0.0015$\,d, with an amplitude that changes in time (see Fig.\ \ref{FIG_timefreq}). Van Kerkwijk et al.\ (2010) suggested that this signal could be associated with the spin period of the A-star. Given our spectroscopic value for $v\sin i$ and the radius determination from our light curve analysis, we expect the spin period of the A-star to be {$0.72\pm0.06\,$d}, which differs by $2\sigma$ from the detected periodicity.

We detected additional variability of similar amplitude in the long cadence Q2 data at frequencies between $15$ and 20\,d$^{-1}$. We could find a similar signal in the observations of KIC\,6889190, which is observed on the same CCD close to \target, and it can be removed succesfully using the cotrending basis vectors technique described in \cite{ChristiansenVan-Cleve2011} (Tom Barclay, priv.\ comm.). This confirms that it is instrumental in nature. 

We did not find significant residual power at the orbital period which proves that the \texttt{LCURVE} model fits the data very well. 

%%%%%%%%%%%%%%%%%%%%%%%%%%%%%%%%%%%%%%%%%%%%
%%  CONCLUSIONS
%%%%%%%%%%%%%%%%%%%%%%%%%%%%%%%%%%%%%%%%%%%%

\section[]{Summary} \label{sec_concl}
We have analysed 212\,d of {\em Kepler} data (Q0, Q1, Q2 and Q3) of the eclipsing binary KOI-74, as well as supporting spectroscopic observations. We modelled the light curve using the \texttt{LCURVE} code, accounting for ellipsoidal modulation, reflection effects, Doppler beaming, R\o mer delay, eclipses and transits. Using Markov Chain Monte Carlo simulations, we determined various system parameters of KOI-74, which are summarised in Table \ref{tab_pars}. We find a lower orbital inclination of {$i=87.0 \pm 0.4^\circ$} compared to the discovery paper and first analysis presented by  \citet{RoweBorucki2010} and adopted by \cite{van-KerkwijkRappaport2010}, $i=88.8 \pm 0.5^\circ$. The difference propagates to our values of other parameters such as the radii. It lowers the mass ratio derived from the ellipsoidal modulation amplitude to {$q=0.052 \pm 0.004$} assuming corotation or {$q=0.047 \pm 0.004$} accounting for a rapidly rotating primary (using $v \sin i = 150 \pm 10\,\kms$ as a prior).

The amplitude of the observed Doppler beaming, which makes the primary become brighter when the star moves in the direction of the {\em Kepler} satellite in its orbit, is in perfect agreement with what is expected from the spectral type and radial velocity amplitude of the star, which we determined to be {$K_1= 15.4 \pm 0.3\,\kms$} from spectroscopy. From the primary's mass of $2.2\pm0.2\,\rm{M}_\odot$ the mass ratio derived from the radial velocity is {$q=0.104\pm0.004$}. 

We also report the first detection of R\o mer delay in a light curve of a compact binary. This delay, which amounts to {$56\pm17\,$s}, is exactly as long as one would expect for a mass ratio of $q\sim0.1$, and is in contradiction with the lower mass ratio derived from the ellipsoidal modulation amplitude. Van Kerkwijk et al. (2010) preferred the mass ratio derived from the Doppler beaming amplitude over the mass ratio derived from the ellipsoidal modulation amplitude. We find that the spectroscopic radial velocity amplitude as well as the R\o mer delay leave no doubt that this was indeed the correct assumption, and that the secondary of KOI-74 is a low mass white dwarf. 

As a result of our lower preferred orbital inclination value, the discrepancy between the mass ratio determined from the ellipsoidal modulation amplitude and the higher mass ratio determined from radial velocity information or R\o mer delay has increased compared to \cite{van-KerkwijkRappaport2010}, and now amounts to a factor 2. Our attempt to account for the effect of the rapid rotation of the primary, increased the discrepancy even further. Our results imply that one has to be very cautious when adopting mass ratio estimates derived from the ellipsoidal modulation amplitude, in particular if there is no firm proof of corotation. We are not aware of any theoretical explanation of the reduction in ellipsoidal modulation as a result of rapid asynchronous rotation.

% -------------------------------------------------------------------
%  Acknowledgements
% -------------------------------------------------------------------

\subsection*{Acknowledgements}
{The authors thank the reviewer, Marten van Kerkwijk, for his detailed and helpful comments on the paper.} SB, PD and CA  acknowledge the KITP staff of UCSB for their warm hospitality
during the research programme ``Asteroseismology in the Space Age". SB acknowledges the travel grant (V446211N) he received from the Fund for Scientific Research of Flanders (FWO), Belgium, for his stay at KITP. This research was supported in part by the National Science Foundation, USA, under Grant No. NSF PHY05-51164.
The HERMES project and team acknowledge support from the Fund for Scientific Research of Flanders (FWO), Belgium, the Research Council of K.U.Leuven, Belgium, the Fonds National Recherches Scientifique (FNRS), Belgium, the Royal Observatory of Belgium, the Observatoire de Gen\`eve, Switzerland and the Th\"uringer Landessternwarte Tautenburg, Germany.
The research leading to these results has received funding from the
European Research Council under the European Community's Seventh Framework Programme
(FP7/2007--2013)/ERC grant agreement n$^\circ$227224 (PROSPERITY), as
well as
from the Research Council of K.U.Leuven grant agreement GOA/2008/04.
During this research TRM, EB, CMC, StP, and BTG were supported under a grant from the UK's 
Science and Technology Facilities Council (STFC, ST/1001719/1).  {The authors acknowledge the Kepler team. Funding for this Discovery mission is provided by NASA's Science Mission Directorate.} For the simulations we used the infrastructure of the VSC -- Flemish Supercomputer Center, funded by the Hercules Foundation and the Flemish Government -- department EWI. This research has made use of  SIMBAD, maintained by the Centre de Donn\'ees astronomiques de Strasbourg; the arXiv preprint service, maintained and operated by the Cornell University Library; and NASA's Astrophysics Data System (ADS).

\bibliographystyle{mn2e}
\bibliography{bibtex}

\label{lastpage}

\end{document}